\documentclass[fleqn,usenatbib]{mnras}
\usepackage{graphicx}
\usepackage{amsmath,amssymb,amsfonts}
\usepackage[cal=cm]{mathalfa}
\usepackage{pgfplots} 
\usepackage{smartdiagram}
\usepackage{enumitem}
\pgfplotsset{compat = newest} 
\DeclareRobustCommand{\VAN}[3]{#2}
\let\VANthebibliography\thebibliography
\def\thebibliography{\DeclareRobustCommand{\VAN}[3]{##3}\VANthebibliography}

\def \la{Lyman alpha}

\newcommand{\be}{\begin{equation}}
\newcommand{\ee}{\end{equation}}

\title[VPFIT addendum]{Addendum: Precision in high resolution absorption line modelling, analytic Voigt derivatives, and optimisation methods}

\author[Lee, Webb, Carswell]{
Chung-Chi Lee$^1$\thanks{lee.chungchi16@gmail.com},
John K. Webb$^1$\thanks{jkw.phys@gmail.com},
Robert F. Carswell$^2$\thanks{rfc@ast.cam.ac.uk}.
\\ \\
$^{1}$Clare Hall, University of Cambridge, Herschel Rd, Cambridge CB3 9AL.\\
$^{2}$Institute of Astronomy, University of Cambridge, Madingley Road, Cambridge, CB3 0HA, UK.
}

\date{}

\begin{document}

\label{firstpage}
\pagerange{\pageref{firstpage}--\pageref{lastpage}}
\maketitle




{\sc vpfit} is a comprehensive and widely used code for the analysis of absorption spectra \citep{ascl:VPFIT2014}. The theoretical methods on which it is based are described in \cite{WebbVPFIT2021}. Section 4.2 of that paper introduces a new enhancement to the code: the replacement of previous finite difference derivative calculations by analytic derivatives of the free parameters required to model a complex of absorption transitions. Derivatives of the Voigt function are needed to compute the gradient vector and Hessian matrix in non-linear least squares methods. This {\it Addendum} is concerned with column density parameters and how one can compute analytic derivatives of the Voigt function for the specific case of {\it summed} column densities, as may be required when, for example, solving for the primordial deuterium to hydrogen ratio at high redshift, averaged over an absorption complex.

Section 4.2 (iii) of \cite{WebbVPFIT2021} provides the analytic Voigt derivatives for the simple situation when all column density parameters are independent. However, in some circumstances it is advantageous to solve for a summed column density parameter i.e. the total column density of a set of absorption components within a complex. The practical method for doing this 
is discussed in detail in \cite{web:VPFIT}, so further details are avoided here. When summed column density parameters are used, the associated analytic Voigt function derivatives are more complicated than those given in Equations (30) of \cite{WebbVPFIT2021} so are provided here. \\

We next explain, using numerical examples, how finite difference derivatives can fail in some cases, when using summed column density parameters, hence motivating the use of analytic expressions in order to guarantee stable algorithms.


\section{Failure of finite difference derivatives in some circumstances}
\label{sec:problem}

To illustrate the numerical instabilities that can arise, consider the following simple example. Suppose we wish to model an absorption complex comprising three adjacent components (i.e. at slightly different redshifts). Suppose further, in this illustrative example, that the model includes two atomic species, C\,{\sc iv} and Si\,{\sc iv}. Following the terminology of the main paper, one species will be a primary, the other secondary. Whilst there is only one primary, there may be multiple secondaries.

Without parameter ties, there are thus three column density parameters for each species. If the {\it total} column density for the complex is a parameter of particular interest, one can assign this quantity to the first of the column density parameters. The advantage of doing so is that the summed column density can be better constrained than the sum of the individual column densities\footnote{Depending on how the other free parameters in the model are arranged, as discussed in \cite{web:VPFIT}}. 

We use a practical example similar to that given in the subsection {\it Common pattern relative ion abundances} in the {\sc vpfit} user guide \citep{web:VPFIT}, with the total C\,{\sc iv} adjusted so that the starting guesses are self-consistent\footnote{This is not a requirement for {\sc vpfit} estimates in starting to find a fit, since the program adjusts the subsidiary  Si\,{\sc iv} values to make them consistent.}:

\begin{verbatim}
C IV 14.75x 2.765821aa 12.69i 0.00 1.00E+00
C IV 13.89x 2.765965ab 17.73j 0.00 1.00E+00
C IV 12.89x 2.765995c  8.31k  0.00 1.00E+00
SiIV 13.96% 2.765821AA 12.69I 0.00 1.00E+00
SiIV 13.10X 2.765965AB 17.73J 0.00 1.00E+00
SiIV 12.10X 2.765995C  8.31K  0.00 1.00E+00
\end{verbatim}

The summed column density of C\,{\sc iv} is $N_t^{\mathrm{C \,\textsc{iv}}} = 10^{14.75}$, the second and third column density components are $N_{1,2}^{\mathrm{C \,\textsc{iv}}} = 10^{13.89}, 10^{12.89}$. The column density of the first component is 
\begin{equation}
    \begin{aligned}
    & N_1^{\mathrm{C \,\textsc{iv}}} = 10^{14.75}-10^{13.89}-10^{12.89} \,, \\ 
    & \rightarrow \quad \log N_1^{\mathrm{C \,\textsc{iv}}} \simeq 14.68 \,.
    \end{aligned}
\end{equation}
The `\%' is a marker, to indicate the start of a new group (Si\,{\sc iv} in this example), such that the first entry in the group is the total Si\,{\sc iv} column density.

The following three examples show the problem if we re-order the individual components and apply two-sided numerical derivative with finite difference derivative $fdd =0.01$.

\vspace{0.1 in}
\noindent {\bf (i) Example 1:} \vspace{0.1 in}\\
Left side:
\begin{verbatim}
C IV 14.75x -> 14.76x
C IV 13.89x
C IV 12.89x
\end{verbatim}
Right side:
\begin{verbatim}
C IV 14.75x -> 14.74x
C IV 13.89x
C IV 12.89x
\end{verbatim}
The column densities of first individual component are
\begin{equation}
    \begin{aligned}
    \mathrm{Left \ side}: \quad &N_1 = 10^{14.68} \rightarrow N_1 = 10^{14.69} \,, \\
    \mathrm{Right \ side}: \quad &N_1 = 10^{14.68} \rightarrow N_1 = 10^{14.66} \,.
    \end{aligned}
\end{equation}

In the above first example, we see that the numerator of the fdd is reasonable and the fdd itself presents no problem. However, now consider a slightly different example.

\vspace{0.1 in}
\noindent {\bf (ii) Example 2:} \vspace{0.1 in}\\
Left side:
\begin{verbatim}
C IV 14.75x -> 14.76x
C IV 14.68x
C IV 12.89x
\end{verbatim}
Right side:
\begin{verbatim}
C IV 14.75x -> 14.74x
C IV 14.68x
C IV 12.89x
\end{verbatim}
The column densities of first individual component are
\begin{equation}
    \begin{aligned}
    \mathrm{Left \ side}: \quad &N_1 = 10^{13.89} \rightarrow N_1 = 10^{13.95} \,, \\
    \mathrm{Right \ side}: \quad &N_1 = 10^{13.89} \rightarrow N_1 = 10^{13.80} \,.
    \end{aligned}
\end{equation}

In this second example, the outcome is poor because the numerator of the fdd is large (0.15) and the derivative loses accuracy.

\vspace{0.1 in}
\noindent {\bf (iii) Example 3:} \vspace{0.1 in} \\
Left side:
\begin{verbatim}
C IV 14.75x -> 14.76x
C IV 13.89x
C IV 14.68x
\end{verbatim}
Right side:
\begin{verbatim}
C IV 14.75x -> 14.74x
C IV 13.89x
C IV 14.68x
\end{verbatim}
The column densities of first individual component are
\begin{equation}
    \begin{aligned}
    \mathrm{Left \ side}: \quad &N_1 = 10^{12.89} \rightarrow N_1 = 10^{13.28} \,, \\
    \mathrm{Right \ side}: \quad &N_1 = 10^{12.89} \rightarrow N_1 = - 10^{12.83} \,.
    \end{aligned}
\end{equation}

In this third example, the result is catastrophic because one side of the fdd interval becomes negative, the fdd interval becomes essentially meaningless, and the numerical derivative fails. Of course the problem has arisen because of the ordering of the three column densities; provided the strongest component is placed first in the grouping, the problem is largely avoided. However, this is not only an undesirable solution, it is sometimes impractical, because even if the parameter guesses are ordered ``sensibly'' at the commencement of the non-linear least squares process, subsequent iterations may reduce the column density of the first component in the group such that the difficulty illustrated in example 3 arises. One can easily find, for example, blends of components where there is no obvious one strong component, relative to others in the system. Therefore, a more robust approach is needed, as discussed next.

\section{The solution to the problem -- analytic derivatives}
\label{sec:solution}


\vspace{0.1 in}
\noindent {\bf (i) Notations:}

\begin{enumerate}[leftmargin=0.5cm]
    \item $N^{p}_t$: summed column density of the primary species
    \item $N^{p}_i$: the column density of the $i^{th}$ component of the primary species
    \item $N^{s}_t$: summed column density of the secondary species
    \item $N^{s}_i$: the column density of the $i^{th}$ component of the secondary species
\end{enumerate}
The column density of the first component of each block is not an independent internal variable within {\sc vpfit}. It is necessary to calculate its derivative at iteration of the minimisation. Note that the default variables in {\sc vpfit} are $\log N^{p}_t$, $\log N^{p}_i$ $(i \neq 1)$, and $\log N^{s}_t$. The relation between the relevant variables are listed as follows.
\begin{equation}
    \begin{aligned}
    & N^{p}_1 = N^{p}_t - \sum_{j=2}^m N^{p}_j \,, \\
    & N^{s}_1 = N^{s}_t - \sum_{j=2}^m N^{s}_j \,, \\
    & \log N^{s}_j = \log N^{p}_j - \log N^{p}_t + \log N^{s}_t \,,
    \end{aligned}
\end{equation}
where $m$ is the total number of component of the leading block (i.e. the primary species).

\vspace{0.1 in}
\noindent {\bf (ii) Case I: Derivative of $N^{p}_t$}

\begin{equation}
\label{eq:deriv1}
    \begin{aligned}
    & \frac{d \log N^{p}_1}{d \log N^{p}_t} = F_1 \,, \\
    & \frac{d \log N^{p}_i}{d \log N^{p}_t} = 0 \quad (i \neq 1) \,, \\
    & \frac{d \log N^{s}_1}{d \log N^{p}_t} = F_2 \,, \\
    & \frac{d \log N^{s}_i}{d \log N^{p}_t} = -1 \quad (i \neq 1) \,,
    \end{aligned}
\end{equation}
where
\begin{equation}
\label{eq:F1F2}
    \begin{aligned}
    & F_1 = \frac{N^{p}_t}{N^{p}_t - \sum_{j=2}^m N^{p}_j} \,, \\
    & F_2 = \frac{N^{s}_t}{N^{s}_t - \sum_{j=2}^m N^{s}_j}-1 = F_1 - 1 \,. \\
    \end{aligned}
\end{equation}

\vspace{0.1 in}
\noindent {\bf (iii) Case II: Derivative of $N^{p}_k$}

\begin{equation}
\label{eq:deriv2}
    \begin{aligned}
    & \frac{d \log N^{p}_1}{d \log N^{p}_k} = - F_3 \,, \\
    & \frac{d \log N^{p}_i}{d \log N^{p}_k} = \delta_{ik} \quad (i \neq 1) \,, \\
    & \frac{d \log N^{s}_1}{d \log N^{p}_k} = - F_3 \,, \\
    & \frac{d \log N^{s}_i}{d \log N^{p}_k} = \delta_{ik} \quad (i \neq 1) \,,
    \end{aligned}
\end{equation}
where $\delta_{ik}$ is the Dirac delta function and
\begin{equation}
    \begin{aligned}
    F_3 = \frac{ N^{p}_k}{N^{p}_t - \sum_{j=2}^m N^{p}_j} \,. \\
    \end{aligned}
\end{equation}

\vspace{0.1 in}
\noindent {\bf (iv) Case III: Derivative of $N^{s}_t$}

\begin{equation}
\label{eq:deriv3}
    \begin{aligned}
    & \frac{d \log N^{p}_1}{d \log N^{s}_t} = 0 \,, \\
    & \frac{d \log N^{p}_i}{d \log N^{s}_t} = 0 \quad (i \neq 1) \,, \\
    & \frac{d \log N^{s}_1}{d \log N^{s}_t} = 1 \,, \\
    & \frac{d \log N^{s}_i}{d \log N^{s}_t} = 1 \quad (i \neq 1) \,,
    \end{aligned}
\end{equation}

\begin{figure}
\centering
\includegraphics[width=0.9\linewidth]{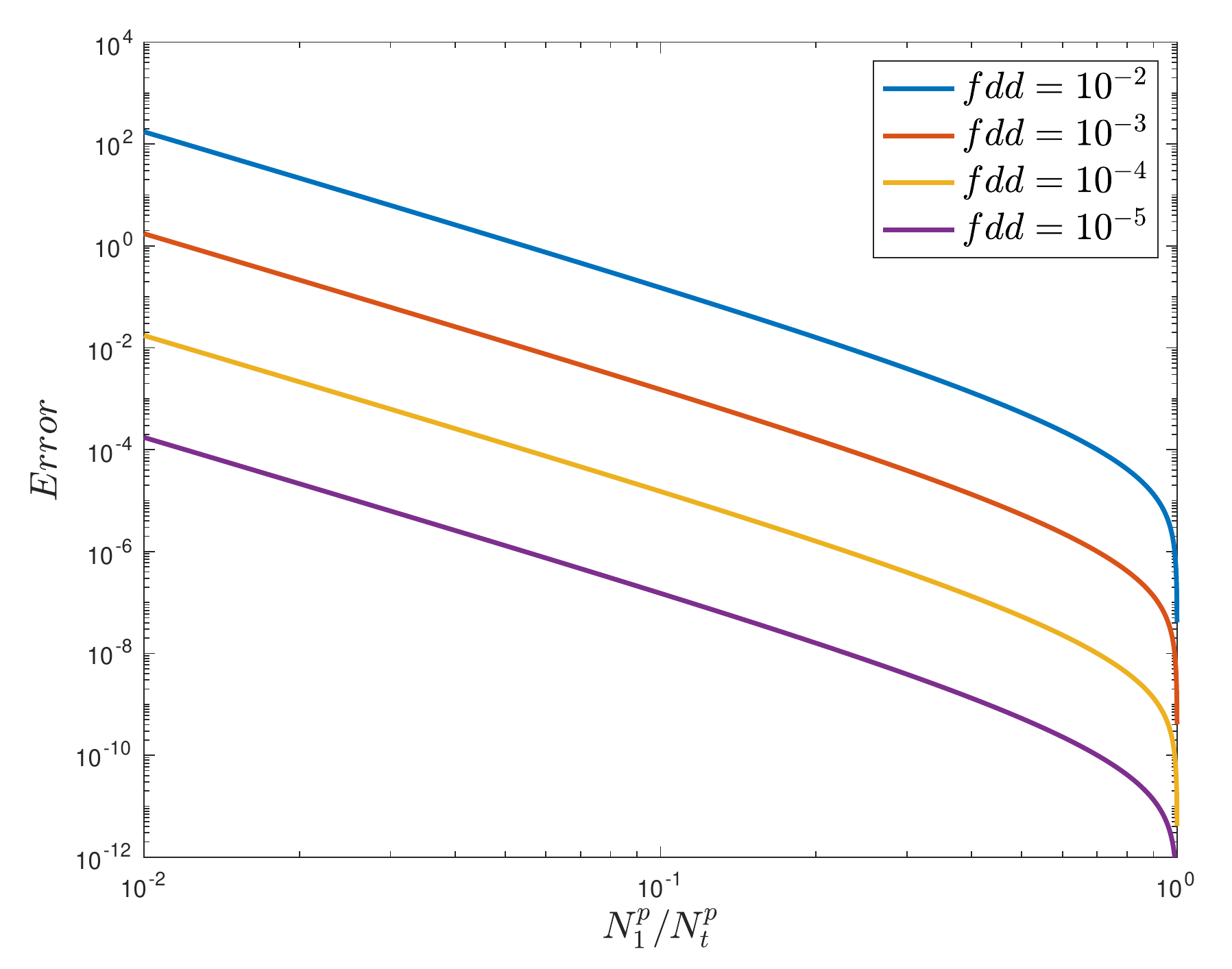}
\caption{The numerical errors go to zero at the right end, which means only the first component has a non-zero column density i.e. only one component is considered.}
\label{fig:err_case1}
\end{figure}

\begin{figure}
\centering
\includegraphics[width=0.9\linewidth]{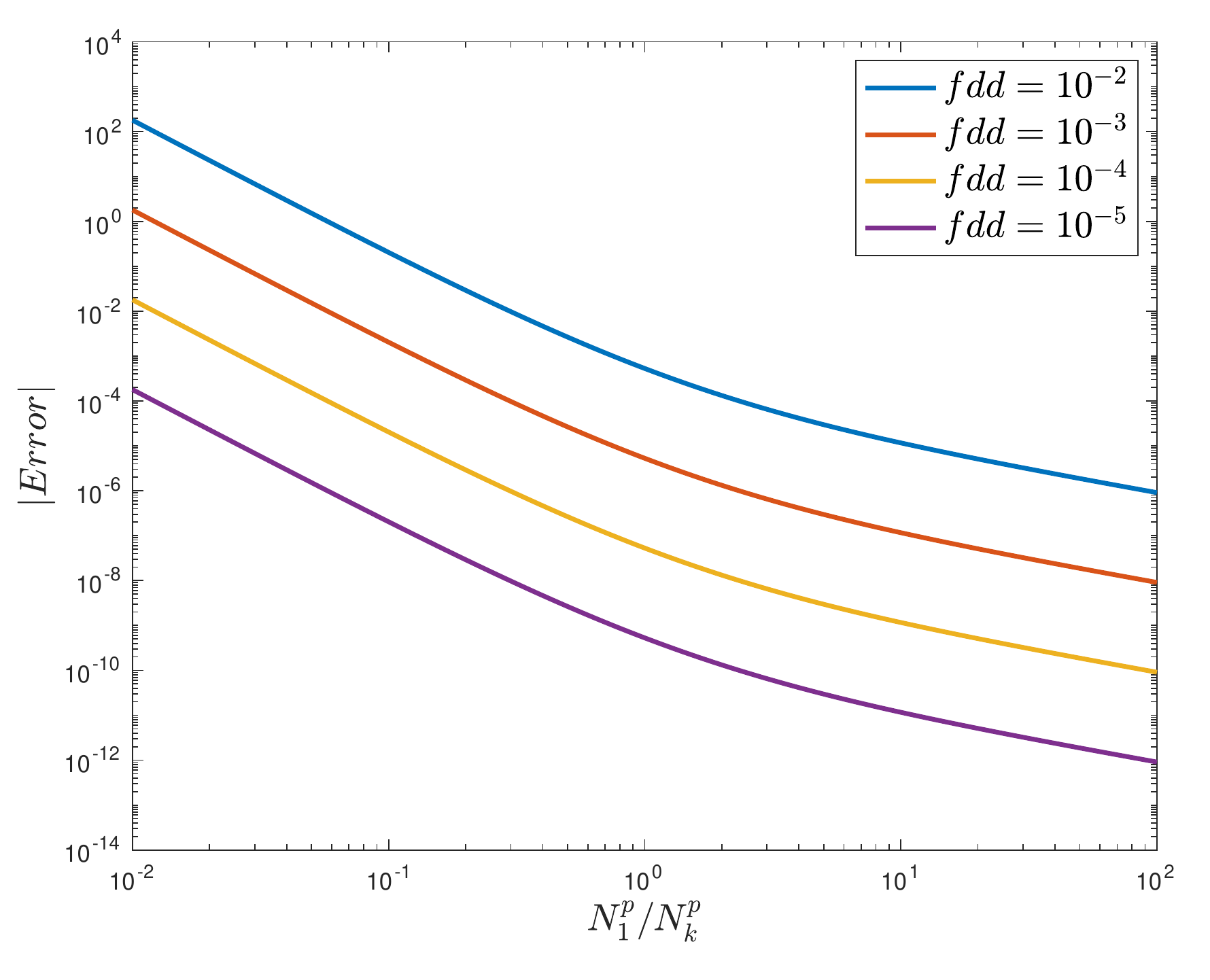}
\caption{Similar to the Fig.~\ref{fig:err_case1}; the numerical error is significant if the first component is weaker than the others.}
\label{fig:err_case2}
\end{figure}

\section{Finite difference derivative numerical errors} \label{sec:numericalerrors}

The now quantify potential fdd numerical errors using the equations given in the previous section, using the first three terms of a Taylor series expansion, which are $e^\Delta = 1+ \Delta + \frac{\Delta^2}{2} \frac{\Delta^3}{6}$ and $\log (1+\Delta) = \Delta - \frac{\Delta^2}{2} + \frac{\Delta^3}{3}$. We ignore cases where the fdd is either 0 or 1 since the fdds do not suffer from the instability described in Section \ref{sec:solution} (see the relevant parts of Eqs.~\eqref{eq:deriv1}, \eqref{eq:deriv2} and \eqref{eq:deriv3}).\\

\vspace{0.1 in}
\noindent {\bf (i) Case I: $\log N^{p}_t \rightarrow \log N^{p}_t \pm \Delta$}

Using a Taylor series expansion we obtain
\begin{equation}
    \begin{aligned}
    \log N^{p}_1 &\rightarrow \log N^{p}_1 + F_1 \cdot \Delta + \left( F_1 - F_1^2 \right) \frac{\ln 10 \cdot \Delta^2}{2} \\ 
    &+ \left( F_1 - 3 F_1^2 + 2F_1^3 \right) \frac{ (\ln 10)^2 \cdot \Delta^3}{6} \,, \\
    \log N^{s}_1 &\rightarrow \log N^{s}_1 + F_2 \cdot \Delta - \left( F_2 - F_2^2 \right) \frac{\ln 10 \cdot \Delta^2}{2} \\ 
    &+ \left( F_2 + 3 F_2^2 + 2F_2^3 \right) \frac{ (\ln 10)^2 \cdot \Delta^3}{6} \,.
    \end{aligned}
\end{equation}
We have the relation from Eq.~\eqref{eq:F1F2},
\begin{equation}
    F_2+3F_2^2+2F_2^3=F_1-3F_1^2+2F_1^3 \,.
\end{equation}
Then, we have
\begin{equation}
    \begin{aligned}
    &\frac{d \log N^{p}_1}{d \log N^{p}_t} \Big\lvert_\mathrm{fdd} - \frac{d \log N^{p}_1}{d \log N^{p}_t} \Big\rvert_\mathrm{analytic} \\ 
    & = \frac{d \log N^{s}_1}{d \log N^{s}_t} \Big\lvert_\mathrm{fdd} - \frac{d \log N^{s}_1}{d \log N^{s}_t} \Big\rvert_\mathrm{analytic} \\ 
    & \simeq \left( F_1 - 3 F_1^2 + 2F_1^3 \right) \frac{ (\ln 10 \cdot \Delta)^2}{6} \,, 
    \end{aligned}
\end{equation}

\vspace{0.1 in}
\noindent {\bf (ii) Case II: $N^{p}_k \rightarrow N^{p}_k \pm \Delta$}

\begin{equation}
\label{eq:n1l-2}
    \begin{aligned}
    \log N^{p}_1 &\rightarrow \log N^{p}_1 - F_3 \cdot \Delta - \left( F_3 + F_3^2 \right) \frac{\ln 10 \cdot \Delta^2}{2} \\ 
    &- \left( F_3 + 3 F_3^2 + 2F_3^3 \right) \frac{ (\ln 10)^2 \cdot \Delta^3}{6} \,.
    \end{aligned}
\end{equation}
The {\sc vpfit} variable, $\log N^{s}_1$, follows the same relation as Eq.~\eqref{eq:n1l-2}. Then, we have
\begin{equation}
    \begin{aligned}
    &\frac{d \log N^{p}_1}{d \log N^{p}_k} \Big\lvert_\mathrm{fdd} - \frac{d \log N^{p}_1}{d \log N^{p}_k} \Big\rvert_\mathrm{analytic} \\ 
    &= \frac{d \log N^{s}_1}{d \log N^{p}_k} \Big\lvert_\mathrm{fdd} - \frac{d \log N^{s}_1}{d \log N^{p}_k} \Big\rvert_\mathrm{analytic} \\ 
    &\simeq - \left( F_3 + 3 F_3^2 + 2F_3^3 \right) \frac{ (\ln 10 \cdot \Delta)^2}{6} \,.
    \end{aligned}
\end{equation}
Figures \ref{fig:err_case1} and \ref{fig:err_case2} illustrate the estimated fdd numerical errors as a function of the relative strength of the first component of the primary species, where
\begin{equation}
    Error = \left( \frac{d \log N^{p}_1}{d \log N^{p}_t} \Big\lvert_\mathrm{fdd} \Big/ \frac{d \log N^{p}_1}{d \log N^{p}_t} \Big\rvert_\mathrm{analytic} \right) -1 \,.
\end{equation}

Both figures illustrate the basic problem i.e. that the fdd numerical error becomes large when the column density of the first component in the group is weak relative to the total column density of that group (Figure \ref{fig:err_case1}) or to a following component of that group (Figure \ref{fig:err_case2}).

\bibliography{vpfit}
\bibliographystyle{mnras}

\bsp
\label{lastpage}
\end{document}